\documentclass[fleqn,twoside]{article}
\usepackage{espcrc2}

\usepackage{graphicx}
\usepackage[figuresright]{rotating}

\def\10{$SO(10)$}
\def\21{SU(2) $\otimes$ U(1) }

\def\422{$SU(4) \otimes SU(2) \otimes SU(2)$}
\def\321{SU(3) $\otimes$ SU(2) $\otimes$ U(1)}
\def \nbb {$\beta\beta_{0\nu}$ }

\def\lsim{\raise0.3ex\hbox{$\;<$\kern-0.75em\raise-1.1ex\hbox{$\sim\;$}}}
\def\gsim{\raise0.3ex\hbox{$\;>$\kern-0.75em\raise-1.1ex\hbox{$\sim\;$}}}

\DeclareMathAlphabet{\mathsc}{OT1}{cmr}{m}{sc}

\def\21{$SU(2) \otimes U(1) $}
\newcommand{\flux}[2][]{\ensuremath{\ifthenelse{\equal{#1}{}}{}{^{#1}\!}\mathit{#2}}}

\newcommand{\CL}   {C.L.}
\newcommand{\dof}  {d.o.f.}
\newcommand{\eVq}  {\mathrm{eV}^2}
\newcommand{\Sol}  {\textsc{sol}}

\newcommand{\Atm}  {\textsc{atm}}

\newcommand{\Dms}  {\Delta m^2_\Sol}
\newcommand{\Dma}  {\Delta m^2_\Atm}

\newcommand{\AddrAHEP}{%
  AHEP Group, Instituto de F\'{\i}sica Corpuscular --
  C.S.I.C./Universitat de Val{\`e}ncia \\
  Edificio Institutos de Paterna, Apt 22085, E--46071 Valencia, Spain}

\title{Neutrino Oscillations and New Physics}

\author{J.~W.~F.~Valle\address[MCSD]\AddrAHEP
  \thanks{Work supported by grant BFM2002-00345, by EC RTN grant
    MRTN-CT-2004-503369.  Latest neutrino oscillation plots taken from
    Ref. \cite{Maltoni:2004ei}.}  }
       
\begin{document}

\begin{abstract}
  I discuss the theoretical background and the status of neutrino
  oscillation parameters from the current worlds' global data sample
  and latest flux calculations.  I give their allowed ranges, best fit
  values and discuss the small parameters $\alpha \equiv \Dms/\Dma$
  and $\sin^2\theta_{13}$, which characterize CP violation in neutrino
  oscillations.
  I mention the significance of \nbb (neutrinoless double beta decay)
  and current expectations in view of oscillation results.
\end{abstract}

\maketitle

\section{INTRODUCTION}

The discovery of neutrino
oscillations~\cite{solarNF,atmNF,reacNu04,accelNu04} marks a turning
point in our understanding of nature and brings neutrino physics to
the center of attention of the particle, nuclear and astrophysics
communities.  The existence of small neutrino masses confirms
theoretical expectations which date back to the early eighties.  They
arise from the dimension-five operator $\ell \ell \phi \phi$ where
$\phi$ is the \21 Higgs doublet and $\ell$ is a lepton
doublet~\cite{Weinberg:1980bf}.  Nothing is known from first
principles about the mechanism that induces this operator, its
associated mass scale or flavour structure.  Its most popular
realization is the seesaw mechanism~\cite{Minkowski:1977sc} which
induces small neutrino masses from the exchange of heavy states, as
expected in unified models. In addition to the standard \21 doublet
Higgs multiplet whose vacuum expectation value (vev) generates gauge
boson and charged fermion masses, the full seesaw model (now called
type-II as opposed to the original terminology in
\cite{schechter:1980gr}), contains Higgses transforming as \21 singlet
and triplet, carrying 2 units of lepton number, and with vevs $v_i$
obeying $v_1 \gg v_2 \gg v_3$ with $v_1 v_3 \sim v_2^2$ (i=1,2,3
correspond to singlet, doublet and triplet, respectively).  The
resulting perturbative description of the seesaw is first given in the
second paper in~\cite{schechter:1980gr}, while its \emph{effective}
model-independent low-energy description involves a 3$\times$6 charged
current lepton mixing matrix which has 24 parameters. These correspond
to 12 mixing angles and 12 CP phases (both Dirac and Majorana-type),
given in~\cite{schechter:1980gr}~(first paper).  Some of these
parameters are involved in leptogenesis~\cite{Fukugita:1986hr}.

Current neutrino oscillation data are well described by the
\emph{simplest} (unitary approximation to the) lepton mixing matrix
neglecting CP violation.  Here I focus mainly on the determination of
neutrino mass and mixing parameters in neutrino oscillation studies, a
currently thriving industry~\cite{industry}, with many new experiments
underway or planned. The interpretation of the data requires good
solar and atmospheric neutrino flux
calculations~\cite{Bahcall:2004fg,Honda:2004yz}, neutrino cross
sections and experimental response functions, and a careful
description of matter effects~\cite{mikheev:1985gs,wolfenstein:1978ue}
in the Sun and the Earth.

In the early eighties it was also argued that, on quite \emph{general}
grounds, well beyond the details of the seesaw mechanism, massive
neutrinos should be Majorana particles~\cite{Wolfenstein:1981rk},
leading to L-violating processes such as \nbb\.  After summarizing the
status of 3-neutrino oscillation parameters I briefly discuss their
impact on future \nbb searches~\cite{Bilenky:2004wn}.

 \section{TWO NEUTRINOS}

\subsection{Solar \& reactor data}
\label{sec:solar-+-kamland}

The solar neutrino data includes the measured rates of the chlorine
experiment at the Homestake mine ($2.56 \pm 0.16 \pm 0.16$~SNU), the
most up-to-date gallium results of SAGE
($66.9~^{+3.9}_{-3.8}~^{+3.6}_{-3.2}$~SNU) and GALLEX/GNO ($69.3 \pm
4.1 \pm 3.6$~SNU), as well as the 1496--day Super-K data in the form
of 44 bins (8 energy bins, 6 of which are further divided into 7
zenith angle bins). The SNO data include the most recent data from the
salt phase in the form of the neutral current (NC), charged current
(CC) and elastic scattering (ES) fluxes, as well as the 2002 spectral
day/night data (17 energy bins for each day and night period).

The analysis methods are described in~\cite{Maltoni:2003da} and
references therein. We use a generalization of the pull approach for
the $\chi^2$ calculation originally suggested in
Ref.~\cite{Fogli:2002pt} in which all systematic uncertainties such as
those of the eight solar neutrino fluxes are included by introducing
new parameters in the fit and adding a penalty function to the
$\chi^2$.  Our generalized method is exact to all orders in the pulls
and covers the case of correlated statistical
errors~\cite{Balantekin:2003jm} as necessary to treat the SNO--salt
experiment.  This is particularly interesting as it allows us to
include the Standard Solar Model $^8$B flux prediction as well as the
SNO NC measurement on the same footing, without pre-selecting a
particular value, as implied by expanding around the predicted value:
the fit itself chooses the best compromise between the SNO NC data and
the SSM prediction.

KamLAND detects reactor anti-neutrinos at the Kamiokande site by the
process $\bar\nu_e + p \to e^+ + n$, where the delayed coincidence of
the prompt energy from the positron and a characteristic gamma from
the neutron capture allows an efficient reduction of backgrounds.
Most of the incident $\bar{\nu}_e$ flux comes from nuclear plants at
distances of $80-350$ km from the detector, far enough to probe the
LMA solution of the solar neutrino problem.
The neutrino energy is related to the prompt energy by $E_\nu =
E_\mathrm{pr} + \Delta - m_e$, where $\Delta$ is the neutron-proton
mass difference and $m_e$ is the positron mass.
For lower energies there is a relevant contribution from geo-neutrino
events to the signal~\cite{Fiorentini:2003ww}. To avoid large
uncertainties associated with the geo-neutrino flux an energy cut at
2.6~MeV prompt energy is applied for the oscillation analysis.

First KamLAND data corresponding to a 162 ton-year exposure gave 54
anti-neutrino events in the final sample, after all cuts, while $86.8
\pm 5.6$ events are predicted for no oscillations with $0.95\pm 0.99$
background events.  The probability that the KamLAND result is
consistent with the no--disappearance hypothesis is less than 0.05\%.
This gave the first evidence for the disappearance of neutrinos
traveling to a detector from a power reactor and the first terrestrial
confirmation of the solar neutrino anomaly.

With a somewhat larger fiducial volume of the detector an exposure
corresponding to 766.3~ton-year (including a reanalysis of the
previous 2002 data) has been given~\cite{reacNu04}.  In total 258
events have been observed, versus $356.2\pm 23.7$ reactor neutrino
events expected in the case of no disappearance and $7.5\pm 1.3$
background events. This leads to a confidence level of 99.995\% for
$\bar\nu_e$ disappearance, and the averaged survival probability is
$0.686 \pm 0.044\mathrm{(stat)} \pm 0.045\mathrm{(syst)}$. Moreover
evidence for spectral distortion is obtained~\cite{reacNu04}.

It is convenient to bin the latest KamLAND data in $1/E_\mathrm{pr}$,
instead of the traditional bins of equal size in $E_\mathrm{pr}$.
Various systematic errors associated to the neutrino fluxes,
backgrounds, reactor fuel composition and individual reactor powers,
small matter effects, and improved $\bar{\nu}_e$ flux parameterization
are included~\cite{Maltoni:2004ei}.  KamLAND data are in beautiful
agreement with the region implied by the LMA solution to the solar
neutrino problem, which in this way has been singled out as the only
viable one in contrast to the previous variety of oscillation
solutions~\cite{Maltoni:2003da,gonzalez-garcia:2000sq}.  However the
stronger evidence for spectral distortion in the recent data leads to
improved information on $\Dms$, substantially reducing the allowed
region of oscillation parameters. From this point of view KamLAND has
played a fundamental role in the resolution of the solar neutrino
problem.

\begin{figure}[htb]
\vglue -.2cm
\includegraphics[height=5cm,width=0.9\linewidth]{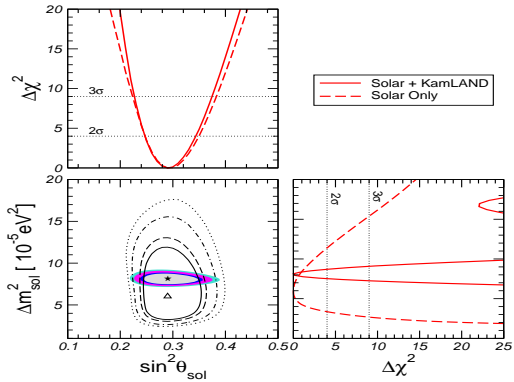}
\vglue -.3cm
\caption{\label{fig:solkam-region} %
  Regions allowed by solar and reactor data at 90\%, 95\%, 99\%, and
  3$\sigma$ \CL\ for 2 \dof. Unshaded regions correspond to solar data
  only.}
\vglue -.5cm
\end{figure}

Assuming CPT one can directly compare the information obtained from
solar neutrino experiments with the KamLAND reactor results.
In Fig.~\ref{fig:solkam-region} we show the allowed regions from the
combined analysis of solar and KamLAND data.  

\subsection{Atmospheric \& accelerator data}
\label{sec:atmospheric-+-k2k}

The zenith angle dependence of the $\mu$-like atmospheric neutrino
data from the Super-K experiment provided the first evidence for
neutrino oscillations in 1998, an effect confirmed also by other
atmospheric neutrino experiments~\cite{atmNF}.  The dip in the $L/E$
distribution of the atmospheric $\nu_\mu$ survival probability seen in
Super-K gives a clearer signature for neutrino oscillations.

\begin{figure}[t] \centering
\vspace{9pt}
    \includegraphics[height=5cm,width=0.9\linewidth]{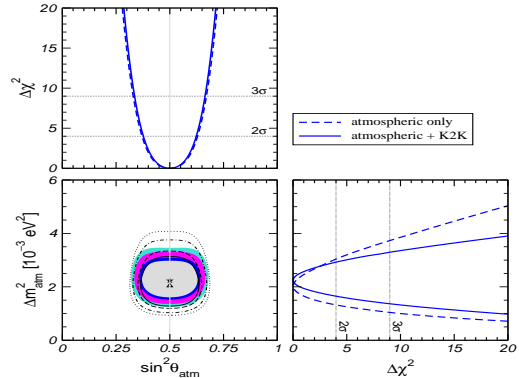}
\vglue -.5cm
    \caption{\label{fig:atm+k2k} %
      $\sin^2\theta_\Atm$--$\Dma$ regions allowed at 90\%, 95\%, 99\%,
      and 3$\sigma$ \CL\ for 2 \dof\ (unshaded regions include
      atmospheric data only). }
\vglue -.8cm
\end{figure}

The analysis summarized below includes the most recent charged-current
atmospheric neutrino data from Super-K, with the $e$-like and
$\mu$-like data samples of sub- and multi-GeV contained events grouped
into 10 zenith-angle bins, with 5 angular bins of stopping muons and
10 through-going bins of up-going muon events.  No information on
$\nu_\tau$ appearance, multi-ring $\mu$ and neutral-current events is
used since an efficient Monte-Carlo simulation of these data would
require more details of the Super-K experiment, in particular of the
way the neutral-current signal is extracted from the data (more
details in Refs.~\cite{Maltoni:2003da,gonzalez-garcia:2000sq}).
In contrast to previous analyses using the Bartol
fluxes~\cite{barr:1989ru}, here we use three--dimensional atmospheric
neutrino fluxes~\cite{Honda:2004yz}.  This way one obtains the regions
of two-flavour $\nu_\mu\to\nu_\tau$ oscillation parameters
$\sin^2\theta_\Atm$ and $\Dma$ shown by the hollow contours in
Fig.~\ref{fig:atm+k2k}. Note that the $\Dma$ values obtained with the
three--dimensional atmospheric neutrino fluxes are lower than obtained
previously~\cite{Maltoni:2003da}, in good agreement with the
results of the Super-K collaboration~\cite{hayato:2003}.

The KEK to Kamioka (K2K) long-baseline neutrino oscillation
experiment~\cite{accelNu04} tests $\nu_\mu$ disappearance in the same
$\Delta m^2$ region as probed by atmospheric neutrinos. The neutrino
beam is produced by a 12~GeV proton beam from the KEK proton
synchrotron, and has 98\% muon neutrinos with 1.3~GeV mean energy. The
beam is controlled by a near detector 300~m away from the proton
target. Information on neutrino oscillations is obtained by comparing
this near detector data with the $\nu_\mu$ content of the beam
observed by the Super-K detector at a distance of 250~km.

The K2K-I data sample gave 56 events in Super-K, whereas
$80.1^{+6.2}_{-5.4}$ were expected for no oscillations. The
probability that the observed flux is explained by a statistical
fluctuation without neutrino oscillations is less than
1\%~\cite{accelNu04}.  K2K-II started in Fall 2002, and released data
at the Neutrino2004 conference~\cite{accelNu04} corresponding to
$4.1\times 10^{19}$ protons on target, comparable to the K2K-I sample.
Altogether K2K-I and K2K-II give 108 events in Super-K, to be compared
with $150.9^{+11.6}_{-10.0}$ expected for no oscillations. Out of the
108 events 56 are so-called single-ring muon events.  This data sample
contains mainly muon events from the quasi-elastic scattering $\nu_\mu
+ p \to \mu + n$, and the reconstructed energy is closely related to
the true neutrino energy.  The K2K collaboration finds that the
observed spectrum is consistent with the one expected for no
oscillation only at a probability of 0.11\%, whereas the spectrum
predicted by the best fit oscillation parameters has a probability of
52\%~\cite{accelNu04}.

The re-analysis of K2K data given in \cite{Maltoni:2004ei} uses the
energy spectrum of the 56 single-ring muon events from K2K-I + K2K-II
(unfortunately not the full K2K data sample of 108 events, for lack of
information outside the K2K collaboration). It is reasonable to fit
the data divided into 15 bins in reconstructed neutrino energy.
One finds that the $\Delta m^2$ indicated by the $\nu_\mu$
disappearance in K2K agrees with atmospheric neutrino results,
providing the first confirmation of oscillations with $\Dma$ from a
man-made neutrino source. However currently K2K gives a weak limit on
the mixing angle due to low statistics.

The shaded regions in Fig.~\ref{fig:atm+k2k} are the allowed
($\sin^2\theta_\Atm$,~$\Dma$) regions that follow from the combined
analysis of K2K and Super-K atmospheric neutrino data.
One sees that, although the determination of $\sin^2\theta_\Atm$ is
completely dominated by atmospheric data, the K2K data start already
to constrain the allowed region of $\Dma$.
Note also that, despite the downward shift of $\Dma$ implied by the
new atmospheric fluxes, the new result is statistically compatible
both with the previous one in~\cite{Maltoni:2003da} and with the value
obtained by the Super-K $L/E$ analysis~\cite{atmNF}.  Note that the
K2K constraint on $\Dma$ from below is important for future
long-baseline experiments, as such experiments are drastically
affected if $\Dma$ lies in the lower part of the 3$\sigma$ range
indicated by the atmospheric data alone.

\section{THREE NEUTRINOS}

The effective leptonic mixing matrix in various gauge theories of
massive neutrinos such as seesaw models was first systematically
studied in~\cite{schechter:1980gr}.  Its simplest unitary form can be
taken as \vglue -.3cm
$$K =  \omega_{23} \omega_{13} \omega_{12}$$
where each factor contains an angle and a phase, 
$$\omega_{13} = \left(\begin{array}{ccccc}
c_{13} & 0 & e^{i \phi_{13}} s_{13} \\
0 & 1 & 0 \\
-e^{-i \phi_{13}} s_{13} & 0 & c_{13} 
\end{array}\right)\,.
$$
This form holds exactly in radiative models of neutrino
mass~\cite{zee:1980ai,babu:1988ki} and approximately in the high-scale
seesaw and models where supersymmetry is the origin of neutrino
mass~\cite{Hirsch:2004he}.  Deviations from unitarity may be
phenomenologically important~\cite{bernabeu:1987gr} in the inverse
seesaw~\cite{mohapatra:1986bd}. Here we stick to the form above.
All three phases in $K$ are physical~\cite{schechter:1981gk}, one
corresponds to the Kobayashi-Maskawa phase of the quarks (Dirac-phase)
and affects neutrino oscillations, while the two Majorana phases show
up in neutrinoless double beta decay and other lepton-number violating
processes~\cite{schechter:1981gk,doi:1981yb}.
Two of the three angles determine solar and atmospheric oscillations,
$\theta_{12} \equiv \theta_\Sol$ and $\theta_{23} \equiv \theta_\Atm$.

Since current neutrino oscillation experiments are not sensitive to CP
violation, we will neglect all phases~(future neutrino factories aim
at probing the effects of the Dirac phase~\cite{Dick:1999ed}). In this
approximation three-neutrino oscillations depend on the three mixing
parameters $\sin^2\theta_{12}, \sin^2\theta_{23}, \sin^2\theta_{13}$
and on the two mass-squared differences $\Dms \equiv \Delta m^2_{21}
\equiv m^2_2 - m^2_1$ and $\Dma \equiv \Delta m^2_{31} \equiv m^2_3 -
m^2_1$ characterizing solar and atmospheric neutrinos.  The hierarchy
$\Dms \ll \Dma$ implies that one can set, to a good approximation,
$\Dms = 0$ in the analysis of atmospheric and K2K data, and $\Dma$ to
infinity in the analysis of solar and KamLAND data.
The relevant neutrino oscillation data in a global three-neutrino
analysis are those of sections \ref{sec:solar-+-kamland} and
\ref{sec:atmospheric-+-k2k} together with the constraints from reactor
experiments~\cite{apollonio:1999ae}.

\begin{figure}[t] \centering
 \includegraphics[width=.96\linewidth,height=5.8cm]{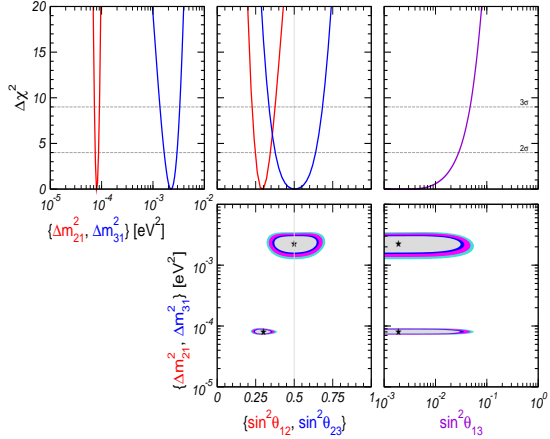}
 \vglue -.5cm
   \caption{\label{fig:global} %
      Three--neutrino regions allowed by the world's oscillation data
      at 90\%, 95\%, 99\%, and 3$\sigma$ \CL\ for 2 \dof}
\vglue -.5cm
\end{figure}

The global three--neutrino oscillation results are summarized in
Fig.~\ref{fig:global} and in Tab.~\ref{tab:summary}. In the upper
panels $\Delta \chi^2$ is shown as a function of the parameters
$\sin^2\theta_{12}, \sin^2\theta_{23}, \sin^2\theta_{13}, \Delta
m^2_{21}, \Delta m^2_{31}$, minimized with respect to the undisplayed
parameters. The lower panels show two-dimensional projections of the
allowed regions in the five-dimensional parameter space. The best fit
values and the allowed 3$\sigma$ ranges of the oscillation parameters
from the global data are summarized in Tab.~\ref{tab:summary}.  This
table gives the current status of the three--flavour neutrino
oscillation parameters.
\begin{table}[t] \centering    \catcode`?=\active \def?{\hphantom{0}}
      \begin{tabular}{|l|c|c|}        \hline        parameter & best
      fit & 3$\sigma$ range         \\        \hline        $\Delta
      m^2_{21}\: [10^{-5}~\eVq]$        & 7.9?? & 7.1--8.9 \\
      $\Delta m^2_{31}\: [10^{-3}~\eVq]$        & 2.2?? &  1.4--3.3 \\
      $\sin^2\theta_{12}$        & 0.30? & 0.23--0.38 \\
      $\sin^2\theta_{23}$        & 0.50? & 0.34--0.68 \\
      $\sin^2\theta_{13}$        & 0.000 & $\leq$ 0.051 \\
      \hline    
\end{tabular}   \vspace{3mm} 
\caption{\label{tab:summary} Three--neutrino oscillation parameters 
from~\cite{Maltoni:2004ei}.}
\vglue -.5cm
\end{table}
  
In a three--neutrino scheme CP violation disappears when two neutrinos
become degenerate~\cite{schechter:1980gr} or when one angle vanishes,
$\theta_{13} \to 0$. Genuine three--flavour effects are associated to
the mass hierarchy parameter $\alpha \equiv \Dms/\Dma$ and the mixing
angle $\theta_{13}$.
\begin{figure}[t] \centering
\vglue -.4cm
    \includegraphics[height=4.2cm,width=.95\linewidth]{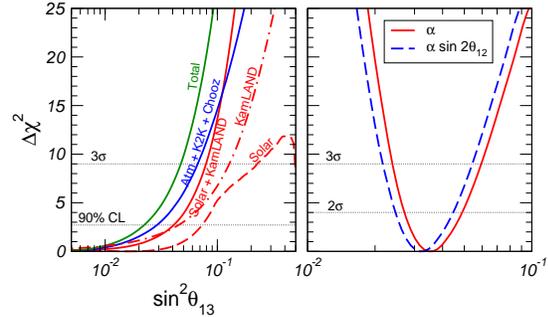}
 \vglue -.4cm
   \caption{\label{fig:alpha}%
      Determination of $\alpha \equiv \Dms / \Dma$ and bound on
      $\sin^2\theta_{13}$ from current  data.}
\vglue -.6cm
\end{figure}
The left panel in Fig.~\ref{fig:alpha} gives the parameter $\alpha$ as
determined from the global $\chi^2$ analysis of \cite{Maltoni:2004ei}.
The figure also gives $\Delta\chi^2$ as a function of the parameter
combination $\alpha \sin 2\theta_{12}$ which, to leading order,
determines the long baseline $\nu_e\to\nu_\mu$ oscillation
probability~\cite{Freund:2001pn,Akhmedov:2004ny}. The last unknown
angle in the three--neutrino leptonic mixing matrix is $\theta_{13}$,
for which only an upper bound exists.
The left panel in Fig.~\ref{fig:alpha} gives $\Delta\chi^2$ as a
function of $\sin^2\theta_{13}$ for different data sample choices.
One finds that the new data from KamLAND have a surprisingly strong
impact on this bound. Before the KamLAND-2004 data the bound on
$\sin^2\theta_{13}$ from global data was dominated by the CHOOZ
reactor experiment, together with the determination of $\Delta
m^2_{31}$ from atmospheric data.  However, with the KamLAND-2004 data
the bound becomes comparable to the reactor bound, and contributes
significantly to the final global bound 0.022~(0.047) at 90\% \CL\ 
(3$\sigma$) for 1 \dof\ This improved $\sin^2\theta_{13}$ bound
follows from the strong spectral distortion found in the 2004
sample~\cite{Maltoni:2004ei}.
\begin{figure}[t] 
\vglue .2cm \centering
    \includegraphics[height=4.5cm,width=.75\linewidth]{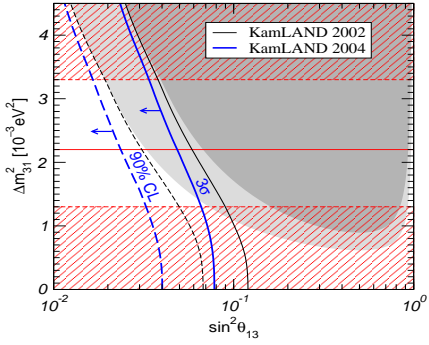}
\vglue -.4cm
    \caption{\label{fig:t13-solar-chooz} Upper bound on
      $\sin^2\theta_{13}$ (1 \dof) from solar+reactor data versus
      $\Dma$. Dashed (solid) curves correspond to 90\% (3$\sigma$)
      \CL\ bounds, thick ones include KamLAND-2004 data, thin ones do
      not.  Light (dark) regions are excluded by CHOOZ at 90\%
      (3$\sigma$) \CL\ The current $\Dma$ best fit value is indicated
      by the horizontal line, hatched regions are excluded by
      atmospheric + K2K data at 3$\sigma$.}  \vglue -.4cm
\end{figure}
Note that, since the reactor bound on $\sin^2\theta_{13}$ quickly
deteriorates as $\Dma$ decreases (see Fig.~\ref{fig:t13-solar-chooz}),
the improvement is especially important at low $\Dma$ values, as
implied by the new three--dimensional atmospheric
fluxes~\cite{Honda:2004yz}.
In Fig.~\ref{fig:t13-solar-chooz} we show the upper bound on
$\sin^2\theta_{13}$ as a function of $\Dma$ from CHOOZ data alone
compared to the bound from an analysis including solar and reactor
neutrino data. One finds that, although for larger $\Dma$ values the
bound on $\sin^2\theta_{13}$ is dominated by CHOOZ, for $\Dma \lsim 2
\times 10^{-3} \eVq$ the solar + KamLAND data start being relevant.
The bounds implied by the 2002 and 2004 KamLAND data are compared in
Fig.~\ref{fig:t13-solar-chooz}. In addition to reactor and accelerator
neutrino oscillation searches, future studies of the day/night effect
in large water Cerenkov solar neutrino experiments like UNO or
Hyper-K~\cite{SKatm04} may give valuable information on
$\sin^2\theta_{13}$~\cite{Akhmedov:2004rq}.

\section{WHAT ELSE}

\begin{figure}[b]
\vglue -.3cm
  \centering
\includegraphics[width=5cm,height=3.2cm]{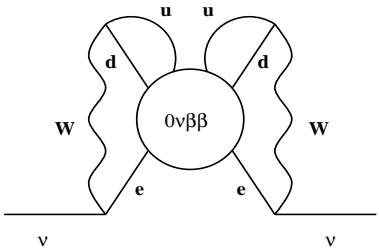}  
\vglue -.3cm
  \caption{Equivalence between \nbb and Majorana mass in gauge theories~\cite{Schechter:1981bd}.}
 \label{fig:bbox} 
\vglue -.8cm
\end{figure}

Neutrino oscillation data are sensitive only to mass differences, not
to the absolute neutrino masses.  Nor do they have any bearing on the
fundamental issue of whether neutrinos are Dirac or Majorana
particles~\cite{schechter:1981gk,doi:1981yb}.  The significance of the
\nbb decay is given by the fact that, in a gauge theory, irrespective
of the mechanism that induces \nbb, it must also produce a Majorana
neutrino mass~\cite{Schechter:1981bd}, as illustrated in Fig.
\ref{fig:bbox}.
Although quantitative implications of this ``black-box'' argument are
model-dependent, any ``natural'' gauge theory obeys the theorem.

Now that oscillations have been confirmed we know that \nbb must be
induced by the exchange of light Majorana neutrinos. The corresponding
amplitude is sensitive both to the absolute scale of neutrino mass as
well as the two Majorana CP phases in the minimal 3-neutrino mixing
matrix~\cite{schechter:1980gr}.
\begin{figure}[t]
  \centering
\includegraphics[width=.7\linewidth,height=6cm]{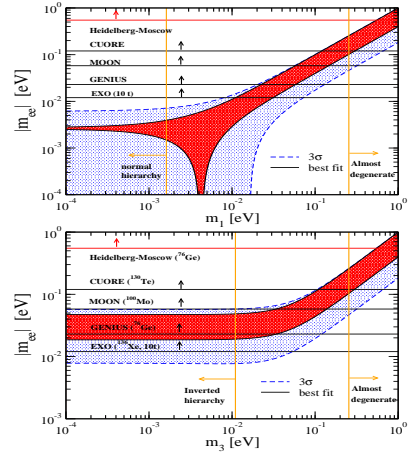}
\vglue -.4cm
  \caption{Neutrino-mass-induced \nbb from current oscillation data 
    versus current and projected experimental sensitivities.}
 \label{fig:nbbfut} 
\vglue -1cm
\end{figure}
Fig. \ref{fig:nbbfut} shows the estimated average mass parameter
characterizing the neutrino exchange contribution to \nbb versus the
lightest neutrino mass.  The upper (lower) panel corresponds to the
cases of normal (inverted) neutrino mass spectra.  The calculation
takes into account the current neutrino oscillation parameters from
\cite{Maltoni:2004ei} and the nuclear matrix elements
of~\cite{Bilenky:2004wn}.
In contrast to the normal hierarchy, where a destructive interference
of neutrino amplitudes is possible, the inverted neutrino mass
hierarchy implies a ``lower'' bound for the \nbb amplitude.
Quasi-degenerate neutrinos~\cite{caldwell:1993kn} such as predicted in
\cite{babu:2002dz}, give the largest \nbb amplitude, as can be seen by
the rising diagonal bands on the right-hand side of the panels. Future
experiments~\cite{Klapdor-Kleingrothaus:1999hk} will provide an
independent confirmation of the present
hint~\cite{Klapdor-Kleingrothaus:2004wj} and push the sensitivity to
inverse hierarchy models.  Complementary information on the absolute
scale of neutrino mass comes from beta decays
searches~\cite{Osipowicz:2001sq} as well as
cosmology~\cite{Hannestad:2004nb}.

In conclusion, we can say that, despite the great progress achieved
recently we are still very far from a ``road map'' to the ultimate
theory of neutrino properties.  We have no idea of the underlying
neutrino mass generation mechanism, its characteristic scale or its
flavor structure. We have still a long way to go and need more data,
especially a confirmation of the LSND and neutrinoless double beta
decay hints.

\end{document}